\documentclass[smallextended]{svjour3}       
\smartqed  

\usepackage[numbers]{natbib}
\usepackage{graphicx}

\begin{document}

\title{Null Geodesics and Wave Front Singularities in the G\"odel Space-time}

\author{Thomas P. Kling         \and
        Kevin Roebuck           \and
        Eric Grotzke
}


\institute{T. Kling \at
              Dept. of Physics, Bridgewater State University,
Bridgewater, MA 02325 \\
              Tel.: +1-508-531-2895\\
              Fax: +1-508-531-1785\\
              \email{tkling@bridgew.edu}           
           \and
           K. Roebuck \at
              Dept. of Physics, Bridgewater State University,
Bridgewater, MA 02325
            \and
            E. Grotzke \at
            Dept. of Physics, Bridgewater State University,
Bridgewater, MA 02325
}

\date{Received: date / Accepted: date}

\maketitle

\begin{abstract}

\noindent We explore wave fronts of null geodesics in the G\"odel metric emitted from point sources both at, and away from, the origin.  For constant time wave fronts emitted by sources away from the origin, we find cusp ridges as well as blue sky metamorphoses where spatially disconnected portions of the wave front appear, connect to the main wave front, and then later break free and vanish.  These blue sky metamorphoses in the constant time wave fronts highlight the non-causal features of the G\"odel metric. We introduce a concept of physical distance along the null geodesics, and show that for wave fronts of constant physical distance, the reorganization of the points making up the wave front leads to the removal of cusp ridges.

\end{abstract}


\maketitle


\section{Introduction} \label{intro:sec}

The G\"odel metric, introduced by Kurt G\"odel in 1949 \cite{godel}, provides a model universe that is an exact solution to the Einstein Field Equations with rotational and non-causal features.  The metric represents  a cosmological solution in which any position in the space-time lies along a stationary world-line about which all other points rotate \cite{hoyle, prasanna}.  The study of G\"odel's metric played an important historical role in clarifying the philosophical underpinnings of general relativity, particularly in regards to Mach's principle \cite{earman}.

While the G\"odel space-time is generally considered to be an unsuitable cosmological solution, it has a number of interesting properties that allow it to serve as a test-bed for physics in other, more relevant and sometimes more complicated, space-times.  One interesting feature of the G\"odel metric is that it allows for closed time-like curves \cite{HE}, a property shared by solutions with space-time singularities but without event horizons.  While this implies that causality is not preserved in the G\"odel universe, the existence of closed time-like curves makes the G\"odel metric similar to the extreme Kerr (spin parameter $a^2>m^2$) solution.  The G\"odel and Kerr metrics also share a common rotational property, manifested in the metric by off-diagonal $dt\,d\phi$ terms.

These interesting properties have led to more recent work introducing a series of G\"odel type metrics and examining their properties, for instance in Griffiths \& Santos (2010) \cite{griffiths}, Carneiro (2002) \cite{carneiro} and Romano \& Goebel (2003) \cite{romano}.  N\'{e}meti et al. (2008) \cite{nemeti} provides excellent visualizations of the light cones, closed time-like geodesics and other features in G\"odel type rotating universes.  Other authors, for example Gleiser et al. (2006) \cite{gleiser} and Nat\'ario (2012) \cite{natario}, have considered properties of the closed time-like geodesics in these space-times, and Slobodov (2008) has shown how changing the topology can be used to remove them \cite{slobodov}.

In this paper, we are interested in exploring the null geodesics of the G\"odel metric and examining wave fronts of the null geodesics emanating from a single point source.  Null geodesics control the causal properties of space-times, and the time evolution of the wave front sweeps out causally connected regions.  Thus, studying the singularities of the wave front and its evolution provides a window into understanding the space-time structure.

The study of wave front singularities is itself an area of mathematical physics research with a long history.  The subject is well described in the general case by V.I. Arnol'd \cite{arnold1}.  Studying wave fronts of light rays in space-times allows one to understand how gravitational lensing arises in a given metric \cite{pettersbook}.  Space-time gravitational lensing, without the typical thin-lens approximations used in applied studies, has been discussed in general in papers \cite{ehlers, perlick}.  In addition, it has been studied in the context of the Schwarzschild metric by Fritelli et al. (2000)\cite{frittelli}, and Rauch and Blandford studied the Kerr metric \cite{rauch}.   The discussion of the formation of wave fronts in the G\"odel space-time allows us to consider gravitational lensing in a different, cosmological style metric.  Our work expands on an initial discussion of wave fronts in G\"odel metrics given by Buser et al. (2013) \cite{buser}.

Of particular interest to us is the impact of rotation and non-causal features on null wave fronts and gravitational lensing.  In standard cosmological space-times, one can define a precise cosmological time, $\tau$, and it makes sense to draw wave fronts of constant $\tau$.  It is less clear in black hole space-times what the wave front slicing should be.  This is complicated further in the extreme Kerr case where the standard Boyer-Lindquist $t$ coordinate has closed time-like curves, as occurs in the G\"odel metric.  One of our purposes in examining null wave fronts from point sources in the G\"odel case is to understand how to draw sensible wave fronts in more physical metrics.

In the first sections of this paper, we derive the equations of motion for null geodesics and indicate how the range in constants of integration are used to sweep out a initial sphere's worth of null geodesics.  We then consider the properties of single null geodesics originating at the origin, $r=0$, and away from the origin.  In section \ref{tfronts:sec}, we show how wave fronts of constant $t$ coordinate evolve for point sources at and away from the origin.  Section \ref{dfronts:sec} introduces a new physical distance, $\delta_p$, in the context of rotating space-times which we believe is a better marker in which to define wave fronts of null geodesics.  Wave fronts of constant $\delta_p$ are drawn, and we discuss the influence this choice makes on the wave front singularities.


\section{Null Geodesics of G\"odel Metric} \label{equations:sec}

The G\"odel space-time metric is given by

\begin{equation} ds^2 = dz^2 -\frac{2}{\omega^2} dt^2 + \frac{2}{\omega^2} dr^2 - \frac{2}{\omega^2} \left( \sinh^4 r - \sinh^2 r \right) d\phi^2 + \frac{4\sqrt{2}}{\omega^2} (\sinh^2 r) dt\,d\phi, \label{metric} \end{equation}

\noindent where $\omega$ is the vorticity of a pressure-free perfect fluid \cite{HE}.  In these coordinates, there is a closed null curve at $r_G=\log(1+\sqrt{2})$ and invariance under changes in $z$.  To find the null geodesics, we consider the Lagrangian

\begin{eqnarray} {\mathcal{L}} & = &  \frac{1}{2} g_{ab} \dot x^a \dot x^b = 0 \nonumber \\ & = &  \frac{1}{2} \dot z^2 - \frac{\dot t^2}{\omega^2} + \frac{\dot r^2}{\omega^2} - \frac{1}{\omega^2} \left( \sinh^4 r - \sinh^2 r \right) \dot \phi^2 + \frac{2\sqrt{2}}{\omega^2} (\sinh^2 r)\,  \dot t \, \dot \phi, \label{lagrangian} \end{eqnarray}

\noindent where the dot indicates a derivative with respect to $s$, an affine parameter.  The condition that the Lagrangian be equal to zero  indicates that we will find null geodesics. We will use that condition to determine the initial conditions and the range of the constants of integration.

Clearly momentum is conserved in the $z$ direction so that

\begin{equation} \dot z = p_z. \end{equation}

\noindent For the $r$ coordinate, the Euler-Lagrange equation yields

\begin{equation} \ddot r = \left( -2 \sinh^3 r \cosh r + \sinh r \cosh r \right) \dot \phi^2 + 2 \sqrt{2} (\sinh r \cosh r) \dot t \, \dot \phi  . \label{reqn} \end{equation}

The situation for $t$ and $\phi$ is more complicated.  The fact that Lagrangian does not depend on the $t$ or $\phi$ coordinates implies the existence of conserved momenta $p_t$ and $p_\phi$.  However, the presence of the $\dot t \dot \phi$ term means that the Euler-Lagrange equations for $t$ and $\phi$ are coupled:

\begin{eqnarray} \frac{-2}{\omega^2} ( \sinh^4 r - \sinh^2 r) \dot \phi + \frac{2\sqrt{2}}{\omega^2} (\sinh^2 r)  \dot t & = &  p_\phi \label{tphi1} \\ \frac{2\sqrt{2}}{\omega^2} (\sinh^2 r) \dot \phi - \frac{2}{\omega^2} \dot t & = & - p_t. \label{tphi2} \end{eqnarray}

\noindent The negative sign in front of $p_t$ in Eq.~\ref{tphi2} allows positive $p_t$ to be associated with generally increasing $t$ values.  Algebraically solving for $\dot t$ and $\dot \phi$ and introducing $v_r = \dot r$, we have five first order ordinary differential equations for light rays in the G\"odel space-time:

\begin{eqnarray} \dot z & = & p_z \label{dotz}\\
\dot t & = & \frac{\omega^2}{2} \left( \frac{\sqrt{2} \, p_\phi}{1 + \sinh^2 r} + \frac{ p_t (1-\sinh^2 r) }{ 1+ \sinh^2 r} \right) \equiv f_t(r) \label{ftdef} \\ \dot \phi & = & \frac{\omega^2}{2} \left( \frac{ p_\phi}{\sinh^2 r \, ( 1+ \sinh^2 r) } - \frac{ \sqrt{2} \, p_t}{1 + \sinh^2 r} \right) \equiv f_\phi(r)  \label{fphidef} \\
\dot r & = & v_r \\ \dot v_r & = & \left( -2 \sinh^3 r \cosh r + \sinh r \cosh r \right) f_\phi^2 + 2 \sqrt{2} \sinh r \cosh r \, f_t f_\phi \label{dotvr}  \end{eqnarray}

\noindent In general, we can integrate null geodesics assuming that $\omega = 1$.  Different values of $\omega$ do not change the overall findings of this paper.  The initial values of $z$ and $t$ can be set to zero with generality because the metric and null geodesics are invariant under shifts in $t$ and $z$.

We will be interested in wave fronts of null geodesics emitted from the origin at $r_o = 0$ as well as wave fronts for general points away from the origin but within the closed null curve at $r_G = \log(1+ \sqrt{2})$. The condition that the geodesics are null geodesics is enforced by setting the Lagrangian to zero at $s=0$, which we enforce by solving Eq.~\ref{lagrangian} for $v_r$ at the initial point.  This leads to

\begin{equation} v_{ro} = \pm \sqrt{ f_{to}^2 - \frac{\omega^2}{2} p_z^2 + (\sinh^4 r_o - \sinh^2 r_o) f_{\phi o}^2 - 2\sqrt{2} (\sinh^2 r_o) f_{\phi o} f_{to} } \label{vro} \end{equation}

\noindent where $r_o$ is the initial radius and $f_{to}$ and $f_{\phi o}$ are the functions defined in Eqs.~\ref{ftdef} and \ref{fphidef} evaluated at the initial radius.  As we will see, requiring the term under the square root to be positive will set limits on the initial momentum $p_z$ and $p_\phi$.

For null geodesics starting at, or passing through, the origin, the angular momentum $p_\phi$ must be zero so that the $\sinh^2 r$ term in the denominator of $f_\phi$ in Eq.~\ref{fphidef} does not lead to an infinity. This is the same general condition that in a flat plane, all geodesics passing through the origin have no angular momentum.  In this case, the initial conditions for null geodesics consist of $r_o = 0$, $t_o = 0$, $z_o = 0$, and then Eq.~\ref{vro} implies

\begin{equation} v_{ro} = + \frac{\omega}{2} \sqrt{\omega^2 p_t^2 - 2 p_z^2}. \label{vroorigin} \end{equation}

\noindent This in turn implies that $p_z^2 < \omega^2 p_t^2 / 2$, or that the $z$ momentum fall in the range $-\sqrt{\omega^2 p_t^2 / 2} < p_z < + \sqrt{\omega^2 p_t^2 / 2} $. The time momentum, $p_t$, is free to take any value, so we can choose $p_t = 1$ and $\omega = 1$ in general.  At the origin, the initial value of $\phi$ is free, and different choices of $\phi_o$ in the range $0<\phi_o < 2\pi$ result in different null geodesics emitted from the origin in a circle's worth of directions spanning the $z=0$ plane.  In practice for rays beginning at the origin, we set $p_t= 1$, pick a $p_z$ value and a $\phi_o$ value, and then use Eq.~\ref{vroorigin} to set the value of $v_r$ at $s=0$.

Null geodesics that begin at $r_o \ne 0$ have different initial conditions and restrictions on the momenta.  The $\phi$ rotational symmetry allows us to take light rays that are emitted from a point $r_o \ne 0$ and $\phi_o = 0$ with $t_o = 0$ and $z_o = 0$.  Then the null geodesic condition is enforced by setting $v_{ro}$ as in Eq.~\ref{vro}.  This implies a more complicated set of conditions on the momentum.  First, in order to ensure that $v_{ro}$ is real, the $z$ momentum is restricted to the range

\begin{equation} |p_z| <  \sqrt{\frac{2}{\omega^2} \left( f_{to}^2 + (\sinh^4 r_o - \sinh^2 r_o) f_{\phi o}^2 - 2\sqrt{2} \sinh^2 r_o f_{to} f_{\phi o} \right) }. \label{pzlim} \end{equation}

\noindent  The range of the momentum $p_\phi$ is then found by requiring the term in the parenthesis of Eq.~\ref{pzlim} to be positive. Multiplying out the functions $f_{t}^2$, $f_\phi^2$ and $f_t f_\phi$ from Eqs.~\ref{ftdef} and \ref{fphidef} results in a quadratic equation for $p_\phi$ whose solution implies that $p_\phi$ be restricted to a range between $p_{\phi-}$ and $p_{\phi+}$ given by

\begin{equation} p_{\phi\pm} =  p_t \, \sinh^2 r_o \left(\sqrt{2} \pm \sqrt{2 + \frac{1-\sinh^2 r_o}{\sinh^2 r_o}} \right). \label{pphilim} \end{equation}

\noindent Again, there are no restriction on $p_t$, so that we set $p_t = 1$ to ensure that at the initial locus of the null wave front, $t$ increases. Choosing a $p_\phi$ within the limits implied by Eq.~\ref{pphilim} and subsequently a value of $p_z$ in the range set by Eq.~\ref{pzlim} spans a sphere's worth of null geodesics originating at a point $(t_o = 0, r_o \ne 0, \phi_o = 0, z_o = 0)$.  Equation \ref{vro} will then set $v_{ro}$ for initially incoming and outgoing light rays.

\section{Numerical Integration} \label{numerica:sec}

While Buser et al. showed that an analytic approach to integrating the null geodesics is possible \cite{buser}, we find it easier to work with numerical integrations.  We implement numerical integration using a modification of the Runge-Kutta-Fehlberg 4-5 adaptive step-size approach outlined in Press et el. (2007) \cite{nrc}.  Adaptive step-size approaches allow one to monitor the accumulated error in the null geodesics.  Because there are no singularities in, or near, the differential equations we are integrating, we are able to keep overall error to less than one part in $10^8$ generically.

We are ultimately interested in wave fronts, with each null geodesic advancing the local region of the wave front. We locate each null geodesic's contribution to the $t=t_1$ wave front by the geodesic's position when in a bin $t_1 + \epsilon$ for small $\epsilon$.  If more than one time step is within the same time bin, we take only the first time step.  Our three-dimensional visualizations are accomplished by separate code that receives data points from the adaptive step-size numerical integration program and then sorts and orders the data according to time (or later constant physical distance). This allows us to examine movies of the moving wave-fronts.

One modification we make to the adaptive step-size approach is to give an {\emph{upper}} limit to the step-size.  This is because in regions where the differential equations are particularly smooth or flat, the natural growth in step-size makes the null geodesics take larger steps than our desired wave front spacing.   In general, the region where the geodesics take naturally smaller steps is when the $r$ coordinate returns close to $r_o$, and the region where the adaptive step-size leads to larger steps is near the outer-most radius - which is actually the area of the greatest physical interest.

\section{Features of Null Geodesics} \label{geodesics:sec}

The constant $z$ momentum implies that the null geodesic path in the $z=0$ plane determines many of the important features of the geodesics.  We first consider rays that pass through the origin, and then we consider null geodesics that do not pass the origin.  The perspective that we will take is that there is an observer at the origin who has arranged sensors throughout the region $r<r_G$ with which she can communicate.

\begin{figure}
\begin{center}
\scalebox{0.8}{\includegraphics{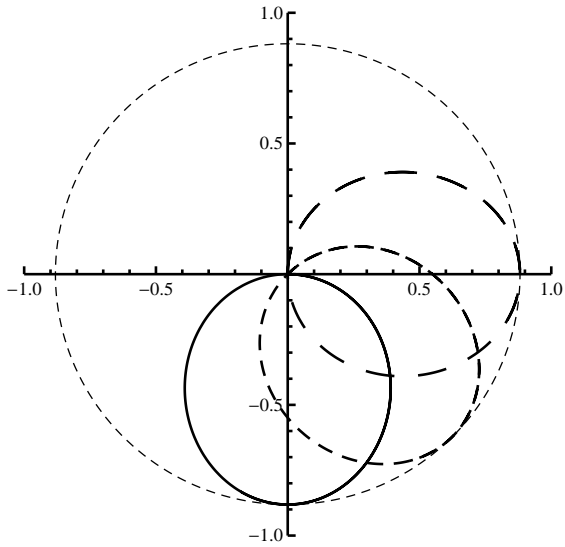}} \caption{\label{rays1:fig}
Three null geodesics with initial $r_o=0$ and $p_z=0$. The solid, short dashed, and long dashed geodesics correspond to initial $\phi$ values of $0$, $\pi/4$ and $\pi/2$.  The trajectories are closed loops in the $z=0$ plane that extend to $r_G = \log(1+\sqrt{2})$ shown as a lightly dashed circle.   } \end{center}\end{figure}

Figure~\ref{rays1:fig} shows the path in the $z=0$ plane of three light rays that originate at the origin.  Each null geodesic has $p_z=0$ and $p_t=1$ for $\omega = 1$.  The initial value of $\phi$ determines the direction of the null geodesics.  The orbits are closed, and they extend out to the $r_G$ limit where they become tangent to the circle of radius $r_G$.

For geodesics passing through the origin, the time coordinate remains timelike along the entire geodesic.  For rays with $p_z=0$, the derivative of $t$ is zero at $r_G$, as one can see from Eq.~\ref{ftdef} with $p_\phi = 0$.  However, for all rays, the $t$ coordinate monotonically increases with the affine parameter $s$.  As we will see, this implies that it will be possible to construct a simply-connected wave front of constant coordinate time, $t$, from different null geodesics originating at the origin with different $\phi$ and $p_z$ values.  Because the null geodesics that pass through the origin can not exit the $r<r_G$ region, we will refer to that region as the observable region in what follows.

\begin{figure}
\begin{center}
\scalebox{0.8}{\includegraphics{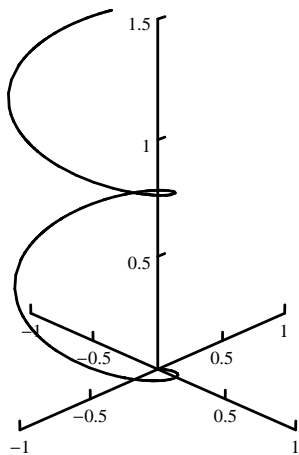}} \caption{\label{3dray:fig}
A null geodesic with $p_z=0.125$ and initial $\phi_o=0$. The negative y axis points towards the lower left and the entire geodesic projects into the $z=0$ plane as a closed curve.} \end{center}\end{figure}

When $p_z$ is not zero, helix shaped null geodesics result as in Fig.~\ref{3dray:fig}.  These geodesics project into the $z=0$ plane as the ellipses, but they do not extend all the way to the $r_G$ limit.  The maximum value of $r$ for a given $p_z$ can be found analytically by solving the equation $\mathcal{L} = 0$ from Eq.~\ref{lagrangian} for $\dot r$ and using the functions $f_t$ and $f_\phi$ defined in Eqs.~\ref{ftdef} and \ref{fphidef}.  With $p_\phi=0$ as is required for rays passing through the origin, the maximum value of $r$ for a given $p_z$ value is the solution to

\begin{equation} \sinh^2 r = \frac{\omega^2 p_t^2 - 2 p_z^2}{\omega^2 p_t^2 + 2 p_z^2}. \end{equation}

\noindent In the $p_z=0$ limit this reduces to $\sinh^2 r = 1$ which is solved by $r=r_G$.

\begin{figure}
\begin{center}
\scalebox{0.8}{\includegraphics{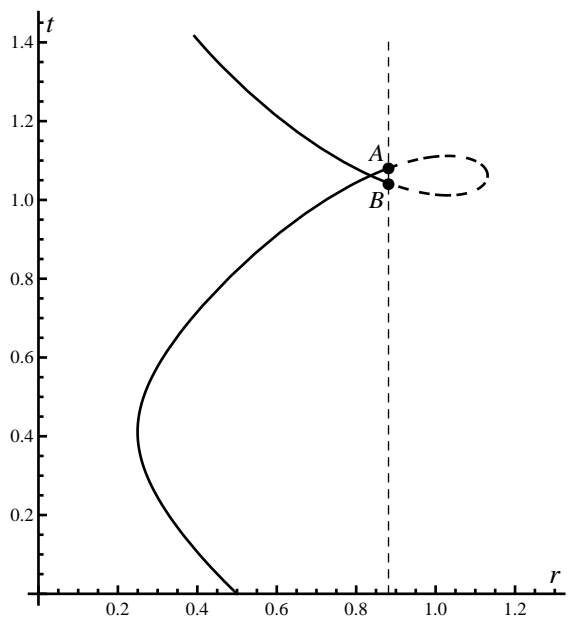}} \caption{\label{rversust:fig}
A plot of the coordinate time versus the radial coordinate for a null geodesic with $p_z=0$ and initial $p_\phi=0.35$ originating at $r_o = 0.5$.  The dotted  line shows  the radius $r_G = \log(1+\sqrt 2)$.  After the null geodesic crosses the $r_G$ radius, the time coordinate loops back so that the time of re-entry $t_B$ is less than the time of exit $t_A$.} \end{center}\end{figure}

The unusual non-causal features of the G\"odel metric manifest themselves when considering null geodesics originating at $r_o \ne 0$.  Figure~\ref{rversust:fig} shows the advancement of the coordinate time, $t$, as a function of the radial coordinate, $r$, for a null geodesic with initial conditions $r_o = 0.5$, $\phi_o = t_o =z_o =0$, $p_t = +1$, $p_\phi = 0.35$, and $p_z=0$.  In terms of the affine parameter, both the $t$ and $r$ coordinates advance smoothly in $s$, but the time coordinate undergoes a brief period where it decreases with the affine parameter, leading to the loop in Fig.~\ref{rversust:fig}. As a result, the null geodesic exits the observable region $r<r_G$ at time $t_A$ and re-enters at an earlier time $t_B$.

In the $z=0$ plane the geodesic is once again a closed curve when we project to points in the $(x,y)$ cartesian plane using

\begin{equation} x = r \cos \phi \quad\quad\quad y = r \sin  \phi. \label{xy} \end{equation}

\begin{figure}
\begin{center}
\scalebox{0.8}{\includegraphics{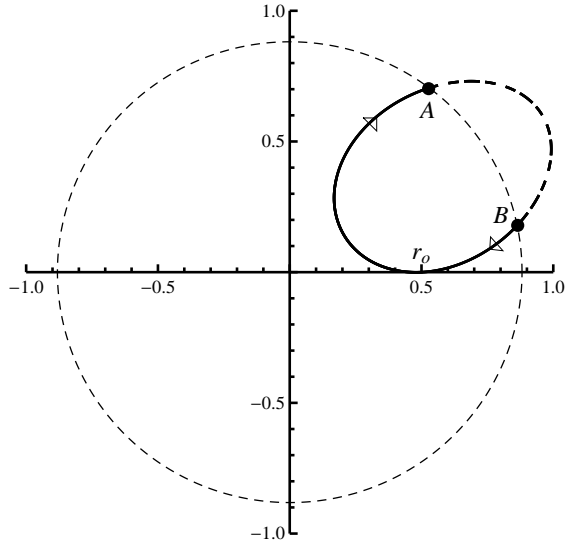}} \caption{\label{rays2:fig}
A null geodesic with $p_z=0$ and initial $p_\phi=0.35$ originating at $r_o = 0.5$.  The dotted circle indicates the radius $r_G = \log(1+\sqrt 2)$.  The null geodesic cycles clockwise until reaching $r_G$ at time $t_A$. The dotted curve shows the continuation of the null geodesics outside the circle of radius $r_G$ falsely using $(x=r\cos\phi, y=r\sin\phi)$.  The null geodesic re-enters the $r<r_G$ region at time $t_B<t_A$. } \end{center}\end{figure}

\noindent However, this null geodesic extends past $r_G$ centered on the origin.  While the values of the $(t, r, \phi, z, v_r)$ coordinates continue to integrate smoothly across this barrier, the meaning of $t$ and $\phi$ coordinates switch with $\phi$ no longer a space-like coordinate.  Thus, the dotted portion of the geodesic in Fig.~\ref{rays2:fig} is not a correct interpretation of the geodesic path through ``space.''  We simply draw it to highlight the continuity of the geodesic path.

Due to the change in how the $t$ coordinate advances in the region $r>r_G$, the null geodesic re-enters the observable region at a time earlier than it left.  This means that for a brief time, to an observer at the origin, the null geodesic would appear to be in two places at the same time.  As a result of the non-monotonic advance of the time coordinate for null geodesics originating at places other than $r_o=0$, the G\"odel universe constant time wave fronts as constructed by an observer at the origin will display unusual features with a disconnected section appearing out of the blue sky along the $r_G$ boundary.

\section{Wave Fronts of Constant Time} \label{tfronts:sec}

A recent paper by Buser et al. (2013) \cite{buser} explores the development of null wave fronts of constant coordinate time in the G\"odel space-time created by a point source at the origin.  We reproduce and expand slightly on their results and then consider light wave fronts from point sources not at the origin. These wave fronts are significantly impacted by the non-causal features of the G\"odel metric.  A wave front of constant $t$ intervals corresponds to wave fronts of constant proper times as observed by a stationary observer at the origin.

\subsection{Wave Fronts Emitted from $r=0$}

\begin{figure}
\begin{center}
\scalebox{0.8}{\includegraphics{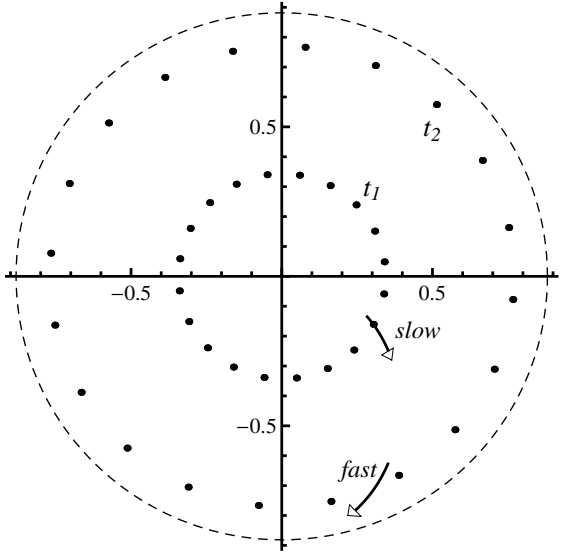}} \caption{\label{planefront:fig}
Dots outline two null geodesic wave fronts of constant coordinate time emitted by a point source at the origin for $p_z =0$.  The inner front is at $t_1$, with the outer front at a later time $t_2$.  The front expands outward with a slow rotation at $t_1$ but rotates quickly in a clockwise direction at $t_2$. } \end{center}\end{figure}

We begin by considering the  wave front in the $z=0$ plane emitted from a point source at the origin.  This wave front is generated by setting $r_o = 0$, $p_z=0$, and $p_\phi=0$ while varying $\phi_o$ in the range zero to $2 \pi$.  At any given time, the wave front itself is circular.  As shown in Fig.~\ref{planefront:fig}, the wave front expands from a point at the origin to the radius $r_G$ and then rebounds, closing back up to a point before expanding back out.

Because of the rotational aspects of the G\"odel space-time, a movie of the wave front expansion would show a rotation of the points in the circle.  Portions of the null geodesic with $\phi_o=0$ are shown in Fig.~\ref{planefront:fig}, with arrows indicating the direction light rays are moving locally. Because the circle $r_G$ is a null curve, the wave front rotates faster as it approaches $r_G$ and generally slows back down (in coordinate time) after rebounding from that limiting circle.

In the full configuration space of $(t, r, \phi, z, v_t, v_r, v_\phi, v_z)$ there is a surface that Ehlers and Newman \cite{ehlers} refers to as the ``lifted wave front.''  The null geodesics are orthogonal to these lifted wave fronts, and the wave fronts show no caustics.  In the G\"odel space-time, there are off-diagonal $dt \, d\phi$ terms that influence the definition of orthogonal.  As a result, the projection of the lifted front into spatial dimensions, in this case the $z=0$ plane, leads to the appearance that the null geodesics are not moving orthogonal to the wave fronts.  One can see this particularly clearly at larger radius in Fig.~\ref{planefront:fig} where the direction of the null geodesic in the projected dimension is not at a visually apparent right angle to the wave front.

\begin{figure}
\begin{center}
\scalebox{0.6}{\includegraphics{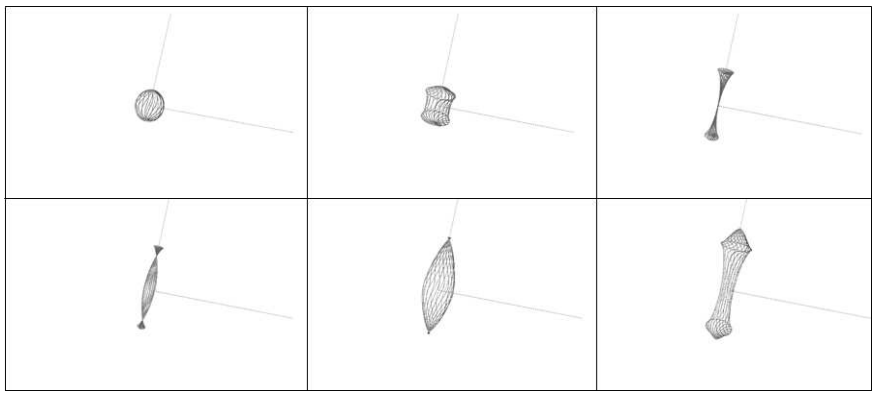}} \caption{\label{originfront:fig}
Null wave fronts of constant coordinate time emitted from a point source at the origin. The $z$ axis points towards the upper right corner, and the viewpoint is fixed in all frames.  An initial spherical front develops a set of cusp ridges as the wave front stretches vertically and oscillates radially. The wave front cycles between the second and sixth frame as time advances.} \end{center}\end{figure}

Figure~\ref{originfront:fig} shows six plots at successive times of the wave front in three space.  The wave front begins as a local sphere, but because the rays that travel more radially, with lower $p_z$ values, reach a limiting cylinder of radius $r_G$ first, they rebound earlier than rays traveling more along the axis of the cylinder.  Individually, the light rays are all following helical motions as in Fig.~\ref{3dray:fig}, so that the entire wave front is rotating.

At any constant $z$ slice, the wave front is circular. Since the individual null geodesics are helixes that all pass repeatedly through points on the $z$ axis, the circle's worth of points with the same $p_z$ value and differing initial $\phi$ values collapse simultaneously along the $z$ axis.  There are always at most two twist points, and the wave front cycles in visual appearance between the second  and final pictures.

\begin{figure}
\begin{center}
\scalebox{0.6}{\includegraphics{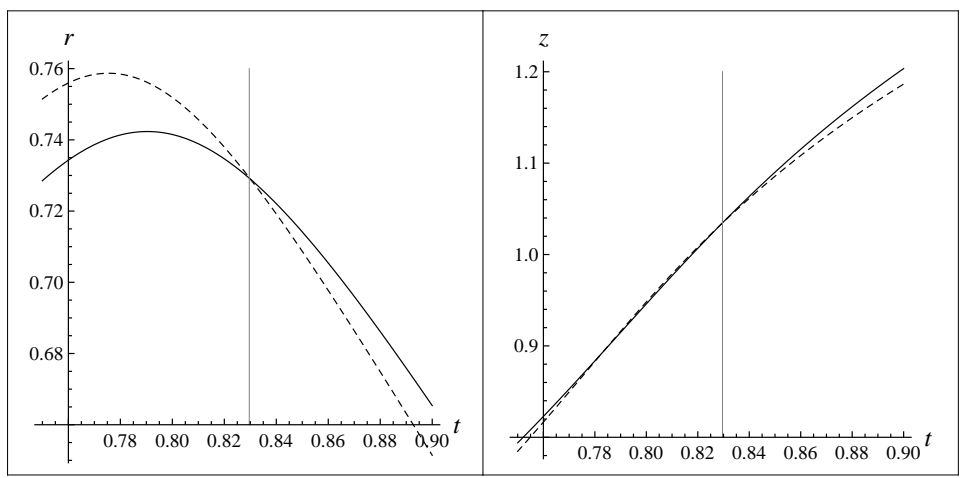}} \caption{\label{cusp:fig}
Plots of the $r$ and $z$ coordinates against $t$ for two neighboring rings emitted from $r_o=0$.  The dotted line corresponds to a ring with $p_z = 0.30$ and the solid line corresponds to $p_z = 0.32$.  The ring with the lower $p_z$ value has a larger peak in its radial extension and rebounds more rapidly in terms of time from this extension, passing through a ring with larger $p_z$ value.  This results in the formation of a cusp ridge.} \end{center}\end{figure}

 Towards the ends of the wave front, we see a cusp ridge.  This form of wave front singularity is common in wave fronts with axial symmetry.  This cusp ridge appears when a circular portion of the wave front with the same $p_z$ value but different initial $\phi$ values catches up with and passes a different circle with a slightly larger $p_z$ value.  Even though $\dot z = dz/ds = p_z$ is constant along the ray, we are plotting wave fronts of constant time, and $dz/dt = p_z/f_t$ varies with the radius of the ring.  Figure~\ref{cusp:fig} plots the $z$ and $r$ coordinates against the $t$ coordinate for two rings with close $p_z$ values.  We see that a ring with a lower $p_z$ value temporarily over-takes one with a higher $p_z$ value, achieving a larger $z$ coordinate for a short time.  These rings cross back over at a time when they have the same radius, which is approximately the location of the cusp at that time.

\subsection{Wave Fronts Emitted from $r\ne 0$}

We now consider light-like wave fronts emitted by a point source away from the origin in the $z=0$ plane.  Our perspective is that there is an observer at the origin who has placed throughout the cylinder given by $r<r_G$ detectors with which she can re-construct the wave front.  Ultimately, we will ignore the portions of the wave front that pass outside this cylinder as the observer at the origin can not receive light signals from detectors placed there.

\begin{figure}
\begin{center}
\scalebox{0.6}{\includegraphics{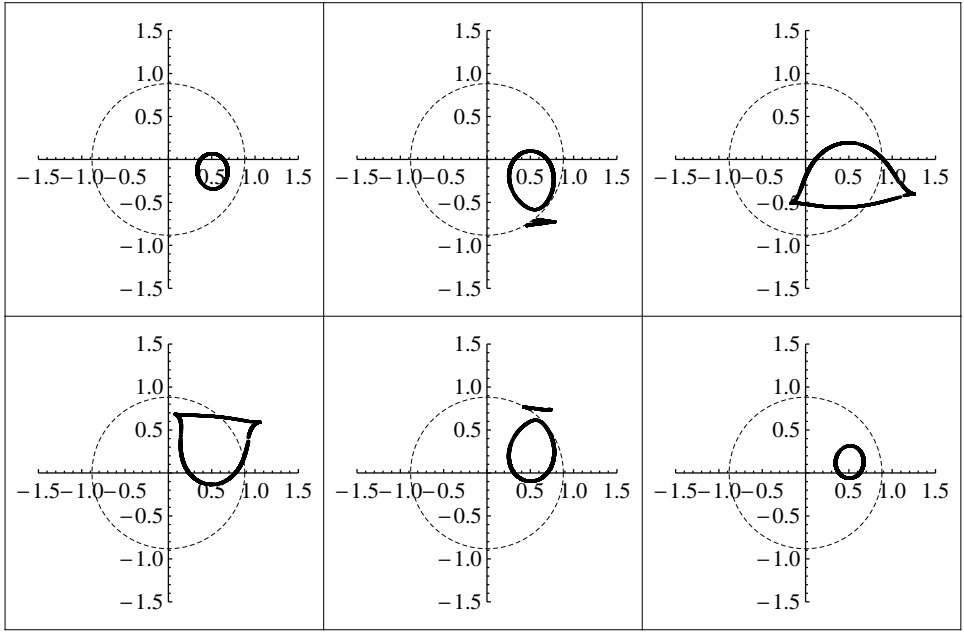}} \caption{\label{planefront2:fig}
Null wave fronts of constant coordinate time emitted by a point source at $r_o=0.5$ at six successive coordinate times. Due to the non-causal nature of the time coordinate, we see that the wavefront undergoes a bifurcation where a new portion appears outside the region observable from the origin and then connects up to the main part of the wavefront. A section later breaks off, shrinks and disappears outside $r_G$.} \end{center}\end{figure}

Figure~\ref{planefront2:fig} shows the constant $t$ wave front in the $z=0$ plane associated with a point source at $r_o = 0.5$, $\phi_o=0$, $p_z=0$. Here $p_\phi$ is varied and $v_{ro}$ is allowed to have positive (initially outgoing directions) and negative (initially incoming directions) values to achieve the circle's worth of initial directions.  We continue to plot the spatial position of the wave fronts using Eq.~\ref{xy}. We do this even though ultimately we consider the portion of the wave front outside the radius $r = r_G$, drawn as a dotted circle, to be un-observable.  Nevertheless, this choice allows us to highlight the fact that a blue sky metamorphosis of the constant time wave front occurs outside the observable region.  This portion of the wave front re-enters the observable region and connects to the portion of the wave front that proceeded uniformly away from the initial location.

Because the wave front is not expanding from the origin, we have broken the axial symmetry of the wave front, and new wave front singularities arise.  We see the appearance of a portion of the wave front ``out of thin air'' or ``out of the blue sky.''  This segment of the wave front has a ``sickle'' shape (Arnol'd language) or ``lips'' (Thom's language) \cite{arnold1}.  These lips connect to the main portion of the wave front and two cusp singularities appear, although only one is inside the observable region. Following the wave front in time, a new set of lips breaks free from the main wave front and vanishes, again outside the observable region.

\begin{figure}
\begin{center}
\scalebox{0.6}{\includegraphics{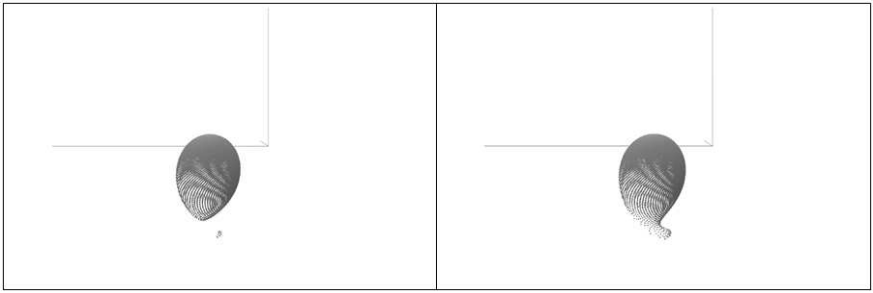}} \caption{\label{bluesky:fig}
On the left, portion of the lips created in a blue sky metamorphosis appear along the cylinder of radius $r_G$ centered at the origin and then combine with the main wave front, as shown at a later time on the right. The positive $z$ axis points into page.} \end{center}\end{figure}

\begin{figure}
\begin{center}
\scalebox{0.6}{\includegraphics{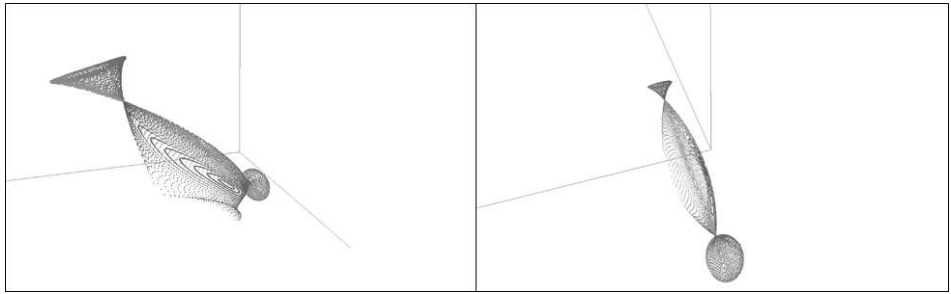}} \caption{\label{r05time:fig}
A top and bottom view at the same time of a wave front of light emitted from a position $r_o = 0.5$ at a late time.  A cusp ridge forms along the bottom side of the wave front where the lips of the blue sky metamorphosis have attached.  The top view shows a cusp ridge on each end of the wave front.} \end{center}\end{figure}

Figure~\ref{bluesky:fig} shows in three dimensions the appearance of a portion of the constant time wave front that has entered the observable region due to the blue sky metamorphosis.  This initially disconnected portion of the wave front merges with the outward expanding wave front slightly later.  In Fig.~\ref{r05time:fig}, we see two views of the wave front at the same time from the top and bottom.  As with the wave fronts emitted from the origin, there is a pair of twist points and a circular cusp ridge near the ends of the wave front.  Unlike the wave front emitted from the origin, the twist points and circular cusp ridges are no longer symmetric with a line parallel to the $z$ axis.  Each circular cusp ridge tilts relative to the $x-y$ plane.  In addition, there is a new cusp ridge along the bottom where the lips from the blue sky metamorphosis have joined the overall wave front.

\section{Defining a Constant Physical Distance} \label{dfronts:sec}

The unusual features of the time coordinate along null geodesics that do not pass through the origin in the G\"odel space-time raise questions about the meaning of constant time wave fronts in general metrics.  In some other metrics, the standard time coordinate may be interpreted in manner that makes more sense.  The clearest examples are the Friedman-Robertson-Walker (FRW) cosmological metrics,

\begin{equation} ds^2 = -dt^2 + a^2(t) dS^2, \label{frw:eqn} \end{equation}

\noindent where there is an easily identified spatial part of the metric, $a^2(t) dS^2$, and a clearly preferred time slicing of the space-time. Alternately, in asymptotically flat space-times, for example the Schwarzschild black hole metric in the typical $(t, r,\theta, \phi)$ coordinates, one could draw constant time wave fronts and interpret these as the wave fronts associated with the time observed by a stationary observer at infinity.  Even though one can make a clear explanation of what the wave fronts mean in this case, these wave fronts have a disadvantage that they appear to ``hang-up'' on the black hole event horizon.

Returning to the FRW case, Eq.~\ref{frw:eqn} implies that the constant time slicing of wave fronts is sensible because it is also a slicing that represents wave fronts of constant physical distance traveled by the light-rays, an insight pointed out by Frittelli and Peters \cite{fp}.  Because light moves at constant speed, time intervals associated with local freely-falling reference frames are truly the correct intervals of time to use to separate the wave fronts.

For this reason, we seek to understand wave fronts of constant physical distance.  In metrics that are diagonal, the constant physical distance is simply integrated along the light ray path using the spatial 3-metric.  The situation is slightly more complicated in the G\"odel metric by the presence of the off-diagonal $dt \, d\phi$ term and by the fact that neither the $t$ nor $\phi$ coordinate has a consistent meaning as a temporal or spatial coordinate at all locations along the light ray. This implies that we should consider two different, but related, coordinate transformations that diagonalize the metric and allow for a clear determination of the spatial 3-metric.

We are working with G\"odel metric in the form of Eq.~\ref{metric}. It is convenient to pull out the time and angular parts of the metric and write them as

\begin{eqnarray} A & = &  \frac{2}{\omega^2} \\ B & = & \frac{4\sqrt{2}}{\omega^2} \sinh^2 r \\ C & = & \frac{2}{\omega^2} \left( \sinh^4 r - \sinh^2 r \right) \end{eqnarray}

\noindent so that the metric has the form

\begin{equation} ds^2 =  -A dt^2 + B dt\, d\phi - C d\phi^2 + \ldots  \label{metricform} \end{equation}

\noindent Inside the observable region, we will introduce a time-like coordinate $\tau$ related to $t$ and $\phi$ by

\begin{equation} d\tau = \sqrt{A} dt - \frac{B}{2\sqrt{A}} d\phi \label{taudef}. \end{equation}

\noindent In terms of this new coordinate, we see that the metric becomes diagonal and there is no mixing of the $\tau$ and $\phi$ coordinates:

\begin{equation} ds^2 = - d\tau^2 + \frac{2}{\omega^2} dr^2 + dz^2 + \left( \frac{B^2}{4 A} - C \right) d\phi^2. \label{metric1} \end{equation}

\noindent The function in front of $d\phi^2$,

\[ \frac{B^2}{4A} - C = \frac{2}{\omega^2} (\sinh^4 r + \sinh^2 r ), \]

\noindent is manifestly positive, so under this coordinate transformation, we can define a physical distance $\delta_p$ along the light ray by integrating

\begin{equation} \delta_p = \int_i^f \, ds \, \sqrt{\frac{2}{\omega^2} \dot r^2 + \dot z^2 + \frac{2}{\omega^2} (\sinh^4 r + \sinh^2 r ) \dot \phi^2 } \label{distance1} \end{equation}

\noindent for $r<r_G$.

Outside the observable region, we introduce a coordinate $\Phi$ defined by

\begin{equation} d\Phi = \sqrt{C} d\phi - \frac{B}{2\sqrt{C}} dt. \label{Phidef} \end{equation}

\noindent In terms of this coordinate, the metric takes the form

\begin{equation} ds^2 = - d\Phi^2 + \left( \frac{B^2}{4C} - A \right) dt^2 + \frac{2}{\omega^2} dr^2 + dz^2 \label{metric2} \end{equation}

\noindent and the term in front of $dt^2$,

\[ \frac{B^2}{4C} - A = \frac{2}{\omega^2} \left( \frac{2 \sinh^2 r}{\sinh^2 r - 1} - 1 \right), \]

\noindent is positive for $r>r_G = \log(1+\sqrt{2})$.  In these coordinates, $d\Phi$ measures a time interval and the spatial 3-metric is identified by the remaining three terms.  Thus for $r>r_G$, the physical distance is integrated from

\begin{equation} \delta_p = \int_i^f \, ds \, \sqrt{\frac{2}{\omega^2} \dot r^2 + \dot z^2 + \frac{2}{\omega^2} \left(\frac{2 \sinh^2 r}{\sinh^2 r - 1} - 1 \right) \dot t^2 } \label{distance2} .
\end{equation}

In practice, we are integrating the null geodesics, Eqs.~\ref{dotz}-\ref{dotvr} numerically.  To keep track of the physical distance along the null geodesics, we simultaneously integrate a sixth ordinary differential equation

\begin{equation} \dot \delta_p = \left\{ \begin{array} {lcr} \sqrt{\frac{2}{\omega^2} v_r^2 + p_z^2 + \frac{2}{\omega^2} (\sinh^4 r + \sinh^2 r ) f_\phi^2 } & \quad\quad &r< r_G \\ ~&~&~\\ \sqrt{\frac{2}{\omega^2} v_r^2 + p_z^2 + \frac{2}{\omega^2} \left(\frac{2 \sinh^2 r}{\sinh^2 - 1} - 1 \right) f_t^2 } & \quad\quad & r>r_G \end{array} \right. \label{distanceode}, \end{equation}

\noindent where we have used the form of the differential equations.  We will use the boundary condition $\delta_p = 0$ at $s=0$. By construction, the physical distance $\delta_p$ will be a monotonically increasing function along the null geodesics, regardless of the origin of the wave fronts. At the boundary $r=r_G$, the physical distance $\delta_p$ is continuous with a discontinuity in the first derivative.

\section{Wave Fronts of Constant Distance}

\begin{figure}
\begin{center}
\scalebox{0.62}{\includegraphics{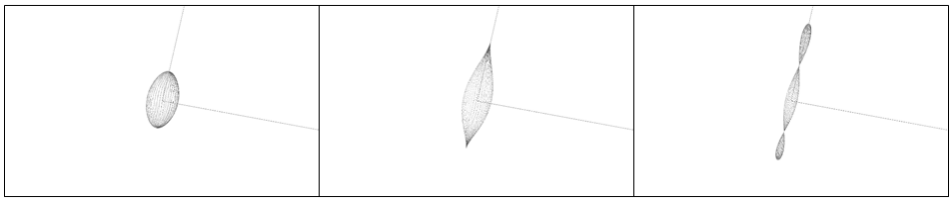}} \caption{\label{r0distance:fig}
Wave fronts of constant physical distance in the G\"odel space time emitted from the origin.  The $z$ axis is oriented up and towards the right to best show the features of the wave front. The cusp ridge present in the constant time wave fronts is missing in the wave fronts of constant physical distance.} \end{center}\end{figure}

We show three successive constant $\delta_p$ wave fronts in Fig.~\ref{r0distance:fig} for a null wave front emitted from the origin.  We see that the general shape of the constant $\delta_p$ wave fronts are similar to those of constant $t$, with a significant difference that the cusp ridges towards the ends of the wave front have vanished.  In terms of progression in $\delta_p$, rings with the same $p_z$ values remain stacked and do not pass each other as we saw that they did in Fig.~\ref{cusp:fig} when we were considering wave fronts of constant time.  The constant $\delta_p$ wave fronts rotate in the same manner as the constant $t$ wave fronts.

\begin{figure}
\begin{center}
\scalebox{0.5}{\includegraphics{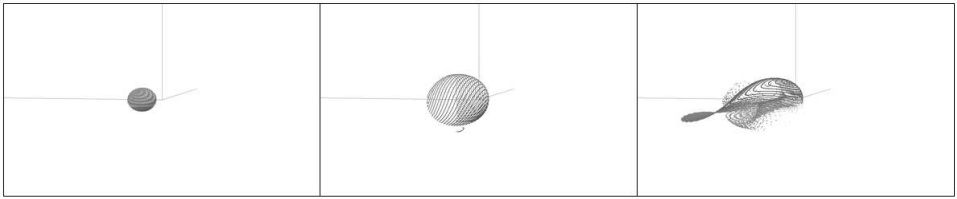}} \caption{\label{r05distance:fig}
Wave fronts of constant physical distance emitted from $r_o = 0.5$.  The cusp ridge vanishes compared to the wave fronts of constant time.  Portions of the wave front continue to re-enter from outside the observable region $r<r_G$, but no longer demonstrate the blue sky metamorphosis.  The $z$ axis is oriented slightly into the page to best show the detached incoming portion of the constant distance wave front. } \end{center}\end{figure}

Three successive wave fronts of constant $\delta_p$ for wave fronts emitted from $r_o = 0.5$ are shown in Fig.~\ref{r05distance:fig}.  Of course, the light rays that make up the wave front continue to leave and re-enter the $r<r_G$ observable region, and so in the central panel of this figure, we do see a disconnected portion of the wave front re-entering the observable region.  This re-entering portion reconnects with the outwardly expanding wave front, and a cusp ridge continues to form where these portions of the wave front connect.  As with the wave front emitted from the origin, the cusp ridges near the end of the expanding wave front have vanished.

If we were to trace the light rays and wave fronts outside the observable region, we would not see a blue sky metamorphosis.  The rays that return to the observable region do so without appearing ``out of nowhere'' or at an earlier time or distance, as was the case with the constant time wave front.  There is simply a portion of the wave front that has exited the observable region and is re-entering after traveling some physical distance.

\section{Discussion}

In this paper, we have examined the wave fronts of null geodesics in the G\"odel space-time emanating both from the origin and from $r_o \ne 0$.  We see that the non-causal features of the G\"odel space-time manifest themselves for wave fronts emanating from a position at $r_o \ne 0$ because null geodesics extend outside the radius $r_G$ and the time coordinate along these geodesics is not monotonically increasing.  In this case, the wave front develops a blue sky metamorphosis where a brand new portion of the wave front, shaped like lips, appears disconnected from the overall wave front.  These results for wave fronts emitted from points not at the origin expand the understanding of null geodesic wave fronts from the work presented in \cite{buser}.

By switching to wave fronts of constant physical distance along the null geodesic, we are able to provide a more physical understanding of wave front evolution in the G\"odel space-time.  For wave fronts emanating from the origin or away from the origin, it is interesting to note that there is a loss of the cusp ridges towards the ends of the wave front.  Cusp ridges are stable features of wave front singularities in the sense that small perturbations of the system do not remove them \cite{arnold1}.  We find it intriguing that the reorganization of points on the wave front from those with constant time to those with constant physical distance is significant enough of a reorganization to remove what is generally considered to be a stable feature under small system perturbations.

For wave fronts of non-origin initial location, we see that the wave fronts lose the ``blue-sky'' style metamorphosis when we transition to constant $\delta_p$ wave fronts.  However, they do not lose the appearance of disconnected sections of the wave front which leave and then re-enter the observable region, reconnecting with the main section of the wave front.    While we consider the constant $\delta_p$ slicing of the wave fronts to be more physical, we note that the appearance of a disconnected region does not vanish.  Nevertheless, the constant coordinate time wave fronts do help explain the ability for the G\"odel metric to support closed time-like curves.

Future work will thoroughly examine the conditions under which switching from a constant coordinate $t$ wave front to a constant physical distance wave front preserves or destroys features of null geodesic wave fronts such as cusps or disconnected regions.  The issue is subtle because when wave fronts are viewed in their entirety, they are inherently non-local, and a global time coordinate may not exist.  Cusps are physical things associated with an increase in magnification: light rays have focussed there.  Hence, the removal of cusps through a global reorganization of the wave front is troubling as it becomes unclear whether light really focussed or not.  Our preliminary examination indicates that the cusps in the G\"odel metric's constant time wave fronts are due to the rebounding off of the $r = r_G$ null surface.  They appear to be similar to the accumulation of light at an event horizon when viewed in constant time slices.  Our initial work indicates that when a constant density spherical dust cloud of radius $r>2m$ is connected to a Schwarzschild exterior, that both the constant time and constant physical distance wave fronts maintain axially symmetric cusp ridges.  While this suggests that the presence of null surfaces is introducing artificial wave front singularities, more analytic and numerical work is needed to confirm this result.

While the G\"odel space-time is interesting in its own right, our primary interest in studying wave fronts of null geodesics in this space-time relate to our interest in a similar problem in the Kerr metric.  There are two relevant similarities. First, both metrics have off-diagonal terms associated with rotational features.  Second, in the case of the extremal Kerr metrics with $a^2>m^2$, the metrics admit closed time-like curves.  Thus, the work done in this paper to derive an expression for physical distances directly carries over to the Kerr case.  It is our intention to examine wave front singularities in the Kerr metric using wave fronts of constant physical distance in a future paper.

\begin{acknowledgements}
KR was supported on this work by a Summer Fellowship from NASA's Massachusetts Space Consortium.  EG thanks the Bridgewater State University Adrian Tinsley Program for
Undergraduate Research for a Summer Grant that enabled his
participation in this project.
\end{acknowledgements}



\end{document}